\documentclass[12pt,showpacs,prd]{revtex4}
\usepackage{epsfig,amssymb,amsfonts}

\makeatletter
\renewcommand{\@makefntext}[1]{\parindent=1em\noindent\hbox to 1.8em{\hss$^{\@thefnmark}$}#1}
\renewcommand{\@footnotemark}{\hbox{\mathsurround=0pt$^{\@thefnmark}$}}
\newcommand{\ftnote}[2]{\footnotemark[#1]\footnotetext[#1]{#2}}
\makeatother
\DeclareMathSymbol{\varGamma}{\mathord}{letters}{"00}

\begin{document}
\title{Confinement and parity doubling in heavy--light mesons}
\author{Yu. S. Kalashnikova}
\affiliation{Institute of Theoretical and Experimental Physics, 117218,\\
B.Cheremushkinskaya 25, Moscow, Russia}
\author{A. V. Nefediev}
\affiliation{Institute of Theoretical and Experimental Physics, 117218,\\
B.Cheremushkinskaya 25, Moscow, Russia}
\author{J. E. F. T. Ribeiro}
\affiliation{ Centro de F\'\i sica das Interac\c c\~oes Fundamentais
(CFIF),Departamento de F\'\i sica, Instituto Superior T\'ecnico, Av.
Rovisco Pais, P-1049-001 Lisbon, Portugal}
\newcommand{\be}{\begin{equation}}
\newcommand{\bea}{\begin{eqnarray}}
\newcommand{\ee}{\end{equation}}
\newcommand{\eea}{\end{eqnarray}}
\newcommand{\ds}{\displaystyle}
\newcommand{\low}[1]{\raisebox{-1mm}{$#1$}}
\newcommand{\loww}[1]{\raisebox{-1.5mm}{$#1$}}
\newcommand{\lmn}{\mathop{\sim}\limits_{n\gg 1}}
\newcommand{\vpint}{\int\makebox[0mm][r]{\bf --\hspace*{0.13cm}}}
\newcommand{\too}{\mathop{\to}\limits_{N_C\to\infty}}
\newcommand{\vp}{\varphi}
\newcommand{\vx}{{\vec x}}
\newcommand{\vy}{{\vec y}}
\newcommand{\vz}{{\vec z}}
\newcommand{\vk}{{\vec k}}
\newcommand{\vq}{{\vec q}}
\newcommand{\vpp}{{\vec p}}
\newcommand{\vn}{{\vec n}}
\newcommand{\vg}{{\vec \gamma}}
\newcommand{\ld}{\lambda}
\newcommand{\cor}{D(\tau,\ld)}

\begin{abstract}
In this paper, we study the chiral symmetry restoration in the
hadronic spectrum in the framework of generalised
Nambu--Jona-Lasinio quark models with instantaneous confining
quark kernels. We investigate a heavy--light quarkonium and derive
its bound--state equation in the form of a Schr{\" o}dingerlike
equation and, after the exact inverse Foldy--Wouthuysen
transformation, in the form of a Diraclike equation. We discuss
the Lorentz nature of confinement for such a system and
demonstrate explicitly the effective chiral symmetry restoration
for highly excited states in the mesonic spectrum. We give an
estimate for the scale of this restoration.
\end{abstract}
\maketitle

\section{Introduction}

Chiral symmetry is spontaneously broken in QCD with light quarks.
This immediately implies the existence of light mesons which
become Goldstone bosons in the chiral limit, with dynamics
described by chiral perturbation theory. Being an effective
low--energy reduction of QCD, this theory does not describe the
property of confinement.

While the property of colour confinement is expected from general
considerations and is supported by lattice calculations, no
effective theory exists which derives confinement directly from
the QCD Lagrangian, and one is left to rely upon models, such as
the constituent quark model, to describe confinement. These models
work surprisingly well in the light--quark sector, reproducing the
spectrum of low--lying mesonic states, with the exception of
ground--state pseudoscalars. So, the conventional wisdom of quark
model practitioners is to assume that constituent quarks are
indeed the correct degrees of freedom in the non-perturbative
domain, with light pseudoscalars being a special case. As there is
no hope to obtain Goldstone bosons within any naive quark model
which does not address the question of QCD vacuum structure, the
sector of light pseudoscalars has to be left outside the scope of
the constituent picture.

Phenomenological successes of constituent models are in fact quite
remarkable, but there is yet another feature of QCD which they
fail to reproduce. Namely, for highly excited hadronic states, we
have quark typical momenta much larger than the scale of chiral
symmetry breaking so that chiral symmetry should be restored
towards the high end of hadronic spectra. The phenomenon
signalling this restoration is parity doubling.

Parity doubling has been discussed for many years in connection
with the baryon spectrum (see, for example, \cite{Glozman}). Quite
recently, the observation of the low--mass $D_s(2317)$ and
$D_s(2460)$ mesons \cite{D} caused a resurgence of interest in the
issue of chiral symmetry restoration \cite{Eichten, Nowak}.
Detailed arguments in favour of parity doubling for higher mesonic
states are given in \cite{Swanson}.

Quite obviously, chiral symmetry restoration is a consequence of
fully relativistic dynamics of light quarks and
chirally--symmetric quark interaction. Besides, it implies the
confining quark interaction, so the models exhibiting parity
doubling should interweave confinement and chiral symmetry
breaking into one single mechanism. These models do exist in the
literature \cite{Orsay,Lisbon}, and the possibility of parity
doubling in such models was briefly discussed in \cite{Orsay2}.
These models are introduced in Chapter II and generic mesonic
bound--state equations are derived for arbitrary kernels. Chapter
III is devoted to the setting up of both the heavy--light bound
state equation and its equivalent Diraclike equation. The
heavy--light meson can be thought of as a kind of hydrogen atom of
soft QCD, with the heavy--quark spin symmetry \cite{IW}, allowing
us to study the dynamics of the light quark ignoring complications
related to the full relativistic two--body problem. Then, for
arbitrary noncompact (that is, confining) quark kernels, it is shown
in Chapter III that these equations necessarily lead to chiral
restoration for high momenta. Finally, we end this chapter with a
numerical study of the heavy--light mesonic spectrum, using, for
the sake of clarity, a simple harmonic oscillator quark kernel
(which allows for a differential mass--gap equation) and explicitly
show the mechanism of chiral restoration at work. We conclude with
Chapter IV.

\section{Mesonic states in potential quark models}

In this section, we briefly give the necessary details of the
model to be considered below. The potential quark model, which we
call the Nambu--Jone-Lasinio (NJL)-type model after the original
paper \cite{NJL} is given by the Hamiltonian
\cite{Orsay,Orsay2,Lisbon}: 
\be 
\hat{H}=\int d^3 x\bar{\psi}(\vec{x},t)\left(-i\vec{\gamma}\cdot
\vec{\bigtriangledown}+m\right)\psi(\vec{x},t)+ \frac12\int d^3
xd^3y\;J^a_\mu(\vec{x},t)K^{ab}_{\mu\nu}(\vec{x}-\vec{y})J^b_\nu(\vec{y},t),
\label{H} 
\ee 
with the quark currents,
$J_{\mu}^a(\vec{x},t)=\bar{\psi}(\vec{x},t)\gamma_\mu\frac{\lambda^a}{2}\psi(\vec{x},t)$,
interacting through an instantaneous kernel, 
\be
K^{ab}_{\mu\nu}(\vec{x}-\vec{y})=g_{\mu 0}g_{\nu 0}\delta^{ab}V_0(|\vec{x}-\vec{y}|). 
\label{KK} 
\ee 
A remarkable feature of the given model is the robustness of the results
against variations of the form and parameters of the confining
potential $V_0(\vx)$. Usually a powerlike form is adopted, 
\be
V_0(|\vec{x}|)=K_0^{\alpha+1}|\vec{x}|^{\alpha}, 
\label{potential}
\ee 
with $0\leqslant\alpha\leqslant 2$. The case of the linearly
rising potential  $\alpha=1$ (see, for example, \cite{linear,Simonov1}) is strongly
supported by hadronic phenomenology, whereas the marginal case of
$\alpha=2$ --- the harmonic oscillator potential --- leads to
simpler, differential, equations and is considered by many authors
\cite{Orsay,Orsay2,Lisbon,pnea} because, despite of its
mathematical simplicity, it already yields a physical picture for
the mechanism of dynamical chiral symmetry breaking not unlike the
one given by linear confinement. 

The quark models of this class fulfill the well-known low energy theorems 
of Gell-Mann, Oakes, and Renner \cite{GOR}, Goldberger and Treiman 
\cite{GT}, Adler selfconsistency zero \cite{ASC}, the 
Weinberg theorem \cite{Wein}, and so on. For an early derivation of the Gell-Mann--Oakes--Renner 
relation, for this class of models, see Ref.~\cite{Orsay2}. For the 
derivation of the Weinberg theorem see Ref.~\cite{EmilCota}. The 
Salpeter equations for these models can be seen in detail in Ref.~\cite{Lisbon}. 
Using this formalism it is possible to give, for 
the class of models embodied in Eq.~(\ref{H}), an 
analytic proof of the Goldberger--Treiman relation \cite{BicudAp}. 
The same formalism can be used to give an analytic proof of all the 
above low-energy theorems. The axial anomaly and the $\pi\gamma\gamma$ coupling constant can
be derived naturally in the framework of the given model as well, following, for example, the lines of the textbook \cite{CL} 
and using the asymptotic freedom and the Ward identity for the dressed axial vertex \cite{BicudAp}
(see also Ref.~\cite{pi2g} for an independent derivation of this relation in the Dyson--Schwinger approach
to the dressed quark propagator).

The standard approach --- valid for
any $\alpha $ --- to the model (\ref{H}) is the Bogoliubov
transformation from bare to dressed quarks, parametrised through
the chiral angle $\vp_p$: 
\be
\psi_{\beta}(\vec{x},t)=\sum_{s=\uparrow,\downarrow}\int\frac{d^3p}{(2\pi)
^3}e^{i\vec{p}\vec{x}} [\hat{b}_{\beta
s}(\vec{p},t)u_s(\vec{p})+\hat{d}_{\beta s}^\dagger(-\vec{p},t)
v_s(-\vec{p})], \label{psi} \ee \be \left\{
\begin{array}{rcl}
u(\vec{p})&=&\frac{1}{\sqrt{2}}\left[\sqrt{1+\sin\vp_p}+
\sqrt{1-\sin\vp_p}\;(\vec{\alpha}\hat{\vec{p}})\right]u(0),\\
v(-\vec{p})&=&\frac{1}{\sqrt{2}}\left[\sqrt{1+\sin\vp_p}-
\sqrt{1-\sin\vp_p}\;(\vec{\alpha}\hat{\vec{p}})\right]v(0),
\end{array}
\right. 
\label{uandv} 
\ee 
where $E_p$ stands for the dispersive
law of the dressed quarks; $\beta$ being the colour index,
$\beta=\overline{1,N_C}$. In what follows the limit $N_C\gg 1$ is
assumed. It is convenient to define the chiral angle varying in
the range $-\frac{\pi}{2}<\vp_p\leqslant\frac{\pi}{2}$, with the
boundary conditions $\vp(0)=\frac{\pi}{2}$, $\vp(p\to\infty)\to
0$.

The Hamiltonian (\ref{H}) normal ordered in the basis (\ref{psi})
can be split into three parts, 
\be 
\hat{H}=E_{\rm vac}+:\hat{H}_2:+:\hat{H}_4:, 
\label{H3} 
\ee 
the first part being
the vacuum energy, and the second and the third parts being
quadratic and quartic in terms of single--quark operators. The
requirement that $:\hat{H}_2:$ should be diagonal --- the
anomalous Bogoliubov term be absent --- leads one to the mass--gap
equation for the chiral angle, 
\be 
A_p\cos\vp_p-B_p\sin\vp_p=0,
\label{mge} 
\ee 
with 
\be
A_p=m+\frac12\int\frac{d^3k}{(2\pi)^3}V(\vec{p}-\vec{k})\sin\vp_k,\quad
B_p=p+\frac12\int \frac{d^3k}{(2\pi)^3}\;(\hat{\vec{p}}\hat{\vec{k}})V(\vec{p}-\vec{k})\cos\vp_k.
\label{AB} 
\ee 
We absorb the fundamental Casimir operator $C_F$
into the definition of the potential,
$V(\vec{p}-\vec{k})=C_FV_0(\vec{p}-\vec{k})$ by rescaling the
parameter $K_0$.

For the chiral angle --- solution to the mass--gap equation
(\ref{mge}) --- the Hamiltonian (\ref{H3}) takes the form: 
\be
\hat{H}=E_{\rm vac}+\sum_\beta\sum_{s=\uparrow,\downarrow}\int
\frac{d^3 p}{(2\pi)^3} E_p[\hat{b}^\dagger_{\beta s}(\vec{p})
\hat{b}_{\beta s}(\vec{p})+\hat{d}^\dagger_{\beta s}(-\vec{p})
\hat{d}_{\beta s}(-\vec{p})]+O\left(\frac{1}{\sqrt{N_C}}\right),
\label{H2diag} 
\ee 
with the corrections coming from the $:\hat{H}_4:$ part. This completes diagonalisation of the Hamiltonian 
(\ref{H}) in the single--particle sector of the theory (the so-called BCS level \cite{BCS}). 
The dressed--quark dispersive law can be built as 
\be 
E_p=A_p\sin\vp_p+B_p\cos\vp_p. 
\label{Ep} 
\ee

It was found in Ref.~\cite{2d} that the 't~Hooft model for
two--dimensional QCD \cite{tHooft} in the axial gauge, which is
identical to the model (\ref{H}) in two dimensions \cite{BG},
admits a further diagonalisation in terms of colourless mesonic
states, which is a step beyond the BCS approximation. This method was generalised, in Ref.~\cite{nr}, to the
class of NJL-type four--dimensional models (\ref{H}). The key idea
was to rewrite the Hamiltonian (\ref{H3}), in the centre--of--mass
frame,
$$
\hat{H}=E_{\rm vac}'+\int \frac{d^3P}{(2\pi)^3}
\hat{\cal H}(\vec{P}),\quad \hat{\cal H}(0)\equiv \hat{\cal H},
\label{HH1}
$$
in terms of compound quark--antiquark operators, 
\be
\hat{M}_{ss'}(\vec{p},\vec{p}')=\frac{\ds 1}{\ds\sqrt{N_C}}\sum_\beta \hat{d}_{\beta
s}(-\vec{p})\hat{b}_{\beta s'}(\vec{p}'), 
\label{operatorsMM} 
\ee
and to perform a second, generalised, Bogoliubov transformation,
\lq\lq dressing" the bare operators (\ref{operatorsMM}) with the
coherent cloud of quark--antiquark pairs. The operators
creating/annihilating the physical mesonic states in the theory
are given by 
\be 
\left\{
\begin{array}{l}
\hat{m}_{n\nu}=\ds\int\frac{p^2dp}{(2\pi)^3}\left[\hat{M}_\nu(p)\vp_{n\nu}^+(p)-
\hat{M}^\dagger_\nu(p)\vp_{n\nu}^-(p)\right]\\
\hat{m}_{n\nu}^\dagger=\ds\int\frac{p^2dp}{(2\pi)^3}\left[\hat{M}_\nu^\dagger
(p)\vp_{n\nu}^+(p)- \hat{M}_\nu(p)\vp_{n\nu}^-(p)\right],
\end{array}
\right. 
\label{mMgen} 
\ee 
where the operators (\ref{operatorsMM}) are expanded first, 
\be 
\hat{M}_{ss'}(\vec{p},\vec{p})=\sum_\nu[\kappa_\nu(\hat{\vec{p}})]_{ss'}\hat{M}_\nu(p), 
\ee 
using an appropriate basis diagonalising the spin--angular structure of
the Hamiltonian. Such a basis is known to be formed by the
$\{nJ^{PC}\}$ states ($n$ being the radial quantum number), for
which we use the shorthand notation $\{n\nu\}$. The centre--of--mass Hamiltonian $\hat{\cal H}$
takes a diagonal form, 
\be 
\hat{\cal H}=\sum_{n,\nu}M_{n\nu}m^\dagger_{n\nu}m_{n\nu}+
O\left(\frac{1}{\sqrt{N_C}}\right), 
\label{hdgen} 
\ee 
provided the mesonic wave functions obey the bound--state equation, 
\be \left\{
\begin{array}{l}
[2E_p-M_{n\nu}]\vp_{n\nu}^+(p)=\ds\int\frac{\ds k^2dk}{\ds (2\pi)^3}
[T^{++}_\nu(p,k)\vp_{n\nu}^+(k)+T^{+-}_\nu(p,k)\vp_{n\nu}^-(k)]\\[0cm]
[2E_p+M_{n\nu}]\vp_{n\nu}^-(p)=\ds\int\frac{\ds k^2dk}{\ds (2\pi)^3}
[T^{-+}_\nu(p,k)\vp_{n\nu}^+(k)+T^{--}_\nu(p,k)\vp_{n\nu}^-(k)],
\end{array}
\right.
\label{bsgen}
\ee
with the orthonormality conditions
\be
\begin{array}{l}
\ds\int\frac{p^2dp}{(2\pi)^3}\left[\vp_{n\nu}^{+}(p)\vp_{m\nu}^{+}(p)-
\vp_{n\nu}^{-}(p)\vp_{m\nu}^{-}(p)\right]=\delta_{nm},\\[3mm]
\ds\int\frac{p^2dp}{(2\pi)^3}\left[\vp_{m\nu}^{+}(p)\vp_{n\nu}^{-}(p)-
\vp_{m\nu}^{-}(p)\vp_{n\nu}^{+}(p)\right]=0,
\end{array}
\ee 
which follow directly from the canonical bosonic commutation
relations of the operators (\ref{mMgen}). This completes the
diagonalisation of the theory (\ref{H}), in terms of the physical
observable degrees of freedom. The bound--state Eq.~(\ref{bsgen}) can be derived alternatively from the
Bethe--Salpeter equation for mesonic states
\cite{Orsay,Orsay2,Lisbon,Lisbon2}. We resort to this method below.
The eigenvalue problem (\ref{bsgen}) is subject to numerical
investigation (see Ref.~\cite{Orsay} and also Ref.~\cite{Lisbon2} where, in particular,
the evolution of the vector--pseudoscalar mass splitting as a 
function of the current quark mass is studied in detail),
and no problem is met when building the mesonic
spectrum for the whole class of theories (\ref{H}).

\section{A Heavy--light system}

\subsection{The bound--state equation}

An important feature of the bound--state Eq.~(\ref{bsgen}) is the
fact that each mesonic state is described by means of a
two--component wave function, $\vp_{n\nu}^\pm(p)$. The physical
meaning of these two components is obvious: $\vp_{n\nu}^+(p)$
describes the forward motion, in time, of the mesonic
quark--antiquark pair and $\vp_{n\nu}^-(p)$ its time backward
motion. Strictly speaking, for a given set of quantum numbers,
four eigenfunctions should be considered
--- two for the eigenvalue $M_{n\nu}$ and two for $-M_{n\nu}$.
Meanwhile, the bound--state Eq.~(\ref{bsgen}) supports the
symmetry \be
\{M_{n\nu},\vp_{n\nu}^\pm(p)\}\leftrightarrow\{-M_{n\nu},\vp_{n\nu}^\mp(p)\},
\label{sym1} 
\ee 
that can be easily verified using its explicit
form. Due to the instantaneous interaction, both particles in the
meson move forward and backward in time {\em in unison}. This
leads us to an immediate conclusion that the heavy--light meson,
with the static--quark Zitterbewegung suppressed by its infinite
mass, can be described with a one--component wave function, so
that 
\be 
\vp_{n\nu}^+(p)\equiv\vp_{n\nu}(p),\quad
\vp_{n\nu}^-(p)=0, \label{sym2} 
\ee 
and the bound--state equation
can be considerably simplified in this case. Although
Eq.~(\ref{bsgen}) is written for the single--flavour theory, its
generalisation for two flavours is trivial yielding for the
heavy--light bound--state equation: 
\be [m_{\bar Q}+E_p-M_{n\nu}]\vp_{n\nu}(p)=\ds\int\frac{\ds k^2dk}{\ds
(2\pi)^3}T^{++}_\nu(p,k)\vp_{n\nu}(k), 
\label{bshl1} 
\ee 
with
$m_{\bar Q}$ being the static--antiquark mass. The amplitude
$T^{++}_\nu(p,k)$ can be built using the Hamiltonian approach
described in the previous section, though we find it more
instructive to derive the heavy--light bound--state
Eq.~(\ref{bshl1}) using an alternative approach, employing the
Bethe--Salpeter equation. To this end, we start from the
homogeneous Bethe--Salpeter equation for the mesonic Salpeter
amplitude $\chi({\vec p};M)$, 
\be 
\chi({\vec p};M)=-i\int\frac{d^4k}{(2\pi)^4}V(\vec{p}-\vec{k})\; \gamma_0
S_q({\vec k},k_0+M/2)\chi({\vec k};M)S_{\bar Q}({\vec k},k_0-M/2)\gamma_0, \label{GenericSal} \ee where \be S_q({\vec
p},p_0)=\frac{\Lambda^{+}({\vec p})\gamma_0}{p_0-E_p+i\epsilon}+\frac{{\Lambda^{-}}({\vec p})\gamma_0}{p_0+E_p-i\epsilon}
\label{Feynman} 
\ee 
is the dressed light--quark propagator. The
projectors are defined as 
\be 
\Lambda^\pm(\vec{p})=T_pP_\pm T_p^\dagger,\quad P_\pm=\frac{1\pm\gamma_0}{2},\quad
T_p=\exp{\left[-\frac12(\vec{\gamma}\hat{\vec{p}})\left(\frac{\pi}{2}-\vp_p\right)\right]}.
\ee 
Since, for the static particle, the chiral angle is simply
$\vp_{\bar Q}(p)=\frac{\pi}{2}$, then the dispersive law and the
projectors become trivial, giving for the heavy--quark propagator:
\be 
S_{\bar Q}({\vec p},p_0)=\frac{P_+\gamma_0}{p_0-m_{\bar Q}+i\epsilon}+ \frac{P_-\gamma_0}{p_0+m_{\bar Q}-i\epsilon}. 
\ee

It is convenient to define a modified vertex, 
\be
\tilde{\varGamma}(\vec{p})=\int\frac{dp_0}{2\pi}S_q(\vec{p},p_0+M/2)\chi(\vec{p};M)S_{\bar
Q}(\vec{p},p_0-M/2), 
\ee 
and to perform its Foldy--Wouthuysen transformation by means of the operator $T_p^\dagger$, 
\be
\varGamma(\vec{p})=T_p^\dagger\tilde{\varGamma}(\vec{p})\hat{1},
\label{Fop} 
\ee 
where, for the sake of transparency, we kept the unity on the r.h.s. for the operator rotating the static
antiquark. The resulting matrix bound--state equation reads: 
\be
(E-E_p)\varGamma(\vec{p})=P_+\left[\int\frac{d^3k}{(2\pi)^3}V(\vec{p}-\vec{k})T_p^\dagger
T_k\varGamma(\vec{k})\right]P_- 
\label{psip}, 
\ee 
where we defined the energy excess over the static--particle mass, $E=M-m_{\bar Q}$. 
The form of the r.h.s. of Eq.~(\ref{psip}) suggests that the
matrix $\varGamma(\vec{p})$ should have the structure 
\be
\varGamma(\vec{p})= \left(
\begin{array}{cc}
0&\psi(\vec{p})\\
0&0
\end{array}
\right)={\psi(\vec{p})\choose 0}_q\otimes(0\;1)_{\bar Q},
\label{qbQ} 
\ee 
where, for future convenience, we split the matrix
into the direct product of the light-- and heavy--particle
components. Then the eigenstate equation for the wave function
$\psi(\vec{p})$ can be derived in the form of a Schr{\" o}dingerlike equation, 
\be
E_p\psi(\vec{p})+\int\frac{d^3k}{(2\pi)^3}V(\vpp-\vk)\left[C_pC_k+
({\vec \sigma}\hat{\vpp})({\vec
\sigma}\hat{\vk})S_pS_k\right]\psi(\vec{k})=E\psi(\vec{p}),
\label{FW4} 
\ee 
with 
\be
C_p=\cos\frac12\left(\frac{\pi}{2}-\vp_p\right),\quad
S_p=\sin\frac12\left(\frac{\pi}{2}-\vp_p\right). 
\label{CSdef} 
\ee
The heavy--light bound--state Eq.~(\ref{FW4}) is the main object
for studies in the present paper. However, it is instructive to
rewrite it in the Diraclike form, which allows one to investigate
the Lorentz nature of confinement in the heavy--light quarkonium.
First, notice that Eq.~(\ref{psip}) is written for the
positive--energy solutions, $M=m_{\bar Q}+E>0$. Similarly, for the
negative--energy solutions, $M=-m_{\bar Q}-E<0$, although the same
Schr{\" o}dingerlike Eq.~(\ref{FW4}) holds, but the matrix
$\varGamma(\vec{p})$ takes a different form, 
\be
\varGamma(\vec{p})= \left(
\begin{array}{cc}
0&0\\
\psi(\vec{p})&0
\end{array}
\right)={0\choose\psi(\vec{p})}_{\bar q}\otimes(1\;0)_Q,
\label{bqQ}
\ee
which can be easily guessed from the symmetry (\ref{sym1})
and the specific property of the heavy--light system (\ref{sym2}).
Thus the Foldy--rotated wave functions of the light component of the
system,
responsible for the positive-- and negative--energy
solutions, can be written, according to Eqs.~(\ref{qbQ}), (\ref{bqQ}), as
$$
\Psi(\vec{p})={\psi(\vec{p})\choose 0},
$$
for positive energies and as
$$
\Psi(\vec{p})={0\choose\psi(\vec{p})},
$$
for negative energies. In view of the passive role played by the
static particle, we shall refer to $\Psi(\vec{p})$ as to the wave
function of the entire heavy--light system. Consequently, the
bound--state Eq.~(\ref{FW4}) can be rewritten for $\Psi(\vec{p})$,
\be 
E_p\Psi(\vec{p})+\frac12\int\frac{d^3k}{(2\pi)^3}V(\vpp-\vk)
\left[T_pT_k^\dagger+T_p^\dagger T_k\right]\Psi(\vec{k})=
\gamma_0E\Psi(\vec{k}). 
\label{int1} 
\ee 
As it follows immediately from Eq.~(\ref{Fop}), the Foldy operator which splits the
positive-- and negative--energy components of this wave function
is $T_p^\dagger$:
$\Psi(\vec{p})=T_p^\dagger\tilde{\Psi}(\vec{p})$. Therefore, the
pre--Foldy bound--state equation for the wave function
$\tilde{\Psi}(\vec{p})$ follows from Eq.~(\ref{int1}) after the
counter--rotation with the operator $T_p$, 
\be
E_pU_p\tilde{\Psi}(\vpp)+\frac12\int\frac{d^3k}{(2\pi)^3}
V(\vpp-\vk)(U_p+U_k)\tilde{\Psi}(\vk)=E\tilde{\Psi}(\vpp),
\label{Se10} 
\ee 
where, for future convenience, we introduced the unitary operator 
\be
U_p=T_p\gamma_0T_p^\dagger=\gamma_0\sin\vp_p+({\vec
\alpha}\hat{\vpp})\cos\vp_p. 
\label{L2} 
\ee

Then, using the mass--gap Eq.~(\ref{mge}), rewritten in the form:
\be 
E_pU_p={\vec \alpha}\vpp+\beta
m+\frac12\int\frac{d^3k}{(2\pi)^3} V(\vpp-\vk)U_k, 
\ee 
we arrive at the heavy--light bound--state equation in the form of a
Diraclike equation for the light quark, in coordinate space, 
\be
({\vec \alpha}\vpp+\beta m)\tilde{\Psi}(\vx)+\int d^3z
\Lambda(\vx,\vz)K(\vx,\vz)\tilde{\Psi}(\vz)=E\tilde{\Psi}(\vx),
\label{DS2} 
\ee 
with 
\be
K(\vx,\vz)=\frac12[V(\vx)+V(\vz)-V(\vx-\vz)] 
\ee 
and 
\be
\Lambda(\vpp,\vq)=(2\pi)^3\delta^{(3)}(\vpp-\vq)U_p.
\label{Lambda0} 
\ee 
Notice that the matrix $\Lambda(\vpp,\vk)$ \cite{Simonov1} can be related to the Green's function as 
\be
\Lambda(\vpp,\vk)=
2i\int\frac{d\omega}{2\pi}S(\omega,\vpp,\vk)\gamma_0,
\label{Lambda} 
\ee 
where we can use either the dressed light--quark propagator $S_q$, defined in Eq.~(\ref{Feynman}),  
$$
S(\omega,\vpp,\vk)=(2\pi)^3\delta^{(3)}(\vpp-\vk)S_q(\vpp,\omega),
$$
or the entire heavy--light Green's function, built using the standard
spectral decomposition, with the help of the solutions of the eigenstate problem (\ref{DS2}),
$$
S(\omega,\vpp,\vk)=\sum_{E_n>0}\frac{\tilde{\Psi}_n(\vpp)\tilde{\Psi}^\dagger_n(\vk)\gamma_0}
{\omega-E_n+i0}+\sum_{E_n<0}\frac{\tilde{\Psi}_n(\vpp)\tilde{\Psi}^\dagger_n(\vk)\gamma_0}
{\omega+E_n-i0}.
$$
This feature should not come as a surprise
since, as mentioned before, the static antiquark decouples from
the system, providing the overall colour neutrality of the bound
state.

The Lorentz nature of confinement in the effective one--particle
Eq.~(\ref{DS2}) follows from the matrix structure of
$\Lambda(\vpp,\vk)$ \cite{Simonov1}, or equivalently, of $U_p$ for
the given values of the relative momentum $p$ involved. Below, we
consider in detail the three regimes which can be identified in
the spectrum of hadrons. Although the heavy--light system was used
before to arrive at the quark--antiquark bound--state equation,
all qualitative conclusions obviously hold for an arbitrary
hadronic system.

For small relative momenta the chiral regime takes place. Chiral
symmetry breaking --- spontaneous or explicit --- plays a
dominating role, the chiral angle being close to $\frac{\pi}{2}$.
As a result, $U_p\approx\gamma_0$ and the effective interquark
interaction \emph{becomes purely scalar (even in the absence of
any microscopic scalar force)}. For light quarks, the
dressed--quark dispersive law $E_p$ differs drastically from the
free--particle form of $\sqrt{\vpp^2+m^2}$, it may even become
negative at small momenta
--- a necessary feature to have the light (massless) chiral pion.
This chiral regime --- where the effective interaction is purely
scalar --- holds below some effective dynamically generated low--energy scale, which we call the BCS scale
$\Lambda_{\rm BCS}$, that is, holds for the mean relative interquark
momentum $p\lesssim\Lambda_{\rm BCS}$. 
The BCS scale gives a measure of chiral symmetry breaking in the low--energy domain of the theory
and it is closely related to the scale of confinement or, equivalently, to $\Lambda_{\rm QCD}$: 
$\Lambda_{\rm BCS}\simeq\Lambda_{\rm QCD}$.
As soon as the mass of the quark--antiquark state grows, as does
the mean relative interquark momentum, the vectorial part of the
interquark interaction becomes more important. Finally, the chiral
angle becomes negligibly small and the interaction acquires a
purely vectorial nature. Thus, we arrive at the chiral symmetry
restoration in the spectrum discussed in the literature
\cite{Swanson,Glozman,Nowak}.

Note, in passing, that the Dirac structure of the matrix
$\Lambda(\vpp,\vk)$ is saturated by the scalar and
space--vectorial part, so there is no room for the time--vectorial
interaction, dangerous from the point of view of the Klein
paradox.

\subsection{Chiral symmetry restoration in the spectrum}

\subsubsection{General remarks}

It is argued in the literature \cite{Orsay2,Swanson,Nowak,Glozman}
that, for highly excited states in the hadronic spectrum, the
chiral symmetry should be restored and the states with opposite
parity should come in doublets. This statement can be proved using
the properties of the bound--state Eq.~(\ref{FW4}). Indeed, let
$\psi(\vec{p})$ be a solution to Eq.~(\ref{FW4}) with the
eigenvalue $E$. Consider 
\be 
\psi'(\vec{p})=({\vec \sigma}\hat{\vpp})\psi(\vec{p}), 
\label{difpar} 
\ee 
which, by construction, possesses an opposite parity to $\psi(\vec{p})$. The
eigenvalue problem for the new wave function
$\psi'(\vec{p})$ can be easily derived by multiplying
Eq.~(\ref{FW4}) by $({\vec \sigma}\hat{\vpp})$ from the left, 
\be
E_p\psi'(\vec{p})+\int\frac{d^3k}{(2\pi)^3}V(\vpp-\vk)\left[S_pS_k+
({\vec \sigma}\hat{\vpp})({\vec \sigma}\hat{\vk})C_pC_k\right]\psi'(\vec{k})=E\psi'(\vec{p}),
\label{FW4pr} 
\ee 
where $C_p$ and $S_p$ are interchanged as
compared to Eq.~(\ref{FW4}). Notice that, for large momenta, one
arrives at the relation $C_p=S_p=\frac{1}{\sqrt{2}}$, which
follows directly from the definition of the functions $C_p$ and
$S_p$ (\ref{CSdef}) and the large--momentum behaviour of the
chiral angle, $\vp_p\mathop{\to}\limits_{p\to\infty} 0$ (see the
first plot at Fig.~1). Therefore, for highly excited states, which
possess a large relative interquark momentum, the bound--state
Eqs.~(\ref{FW4}) and (\ref{FW4pr}) coincide, being simply 
\be
E_p\psi(\vec{p})+\frac12\int\frac{d^3k}{(2\pi)^3}V(\vpp-\vk)\left[1+({\vec \sigma}\hat{\vpp})({\vec
\sigma}\hat{\vk})\right]\psi(\vec{k})=E\psi(\vec{p}),
\label{FW4pr2} 
\ee 
so that the states $\psi(\vec{p})$ and $({\vec
\sigma}\hat{\vpp})\psi(\vec{p})$ become degenerate. Thus, in the
framework of the potential quark model (\ref{H}), one can
investigate the explicit mechanism of the chiral symmetry
restoration high in the hadronic spectrum. For further
convenience, in addition to the BCS scale $\Lambda_{\rm BCS}$
defined above, which can be related, for example, to the chiral
condensate, 
\be
\langle\bar{q}q\rangle=-\frac{3}{\pi^2}\int_0^\infty
dp\;p^2\sin\vp_p\equiv -(\Lambda_{\rm BCS})^3, 
\label{qaq} 
\ee 
we introduce the scale of the symmetry restoration $\Lambda_{\rm
rest}$, to be defined as {\em the scale at which the splitting
within the chiral doublet is much smaller than the BCS scale
$\Lambda_{\rm BCS}$}. Such a definition is natural since,
alternatively, the BCS scale can be defined as the splitting
within a chiral doublet at small values of the relative interquark
momentum, that is, in the region where this splitting is
essentially dictated by chiral symmetry. The numerical estimate
for $\Lambda_{\rm rest}$ in the case of the harmonic confinement
($\alpha =2$) is given below.

\begin{figure}[t]
\centerline{\epsfig{file=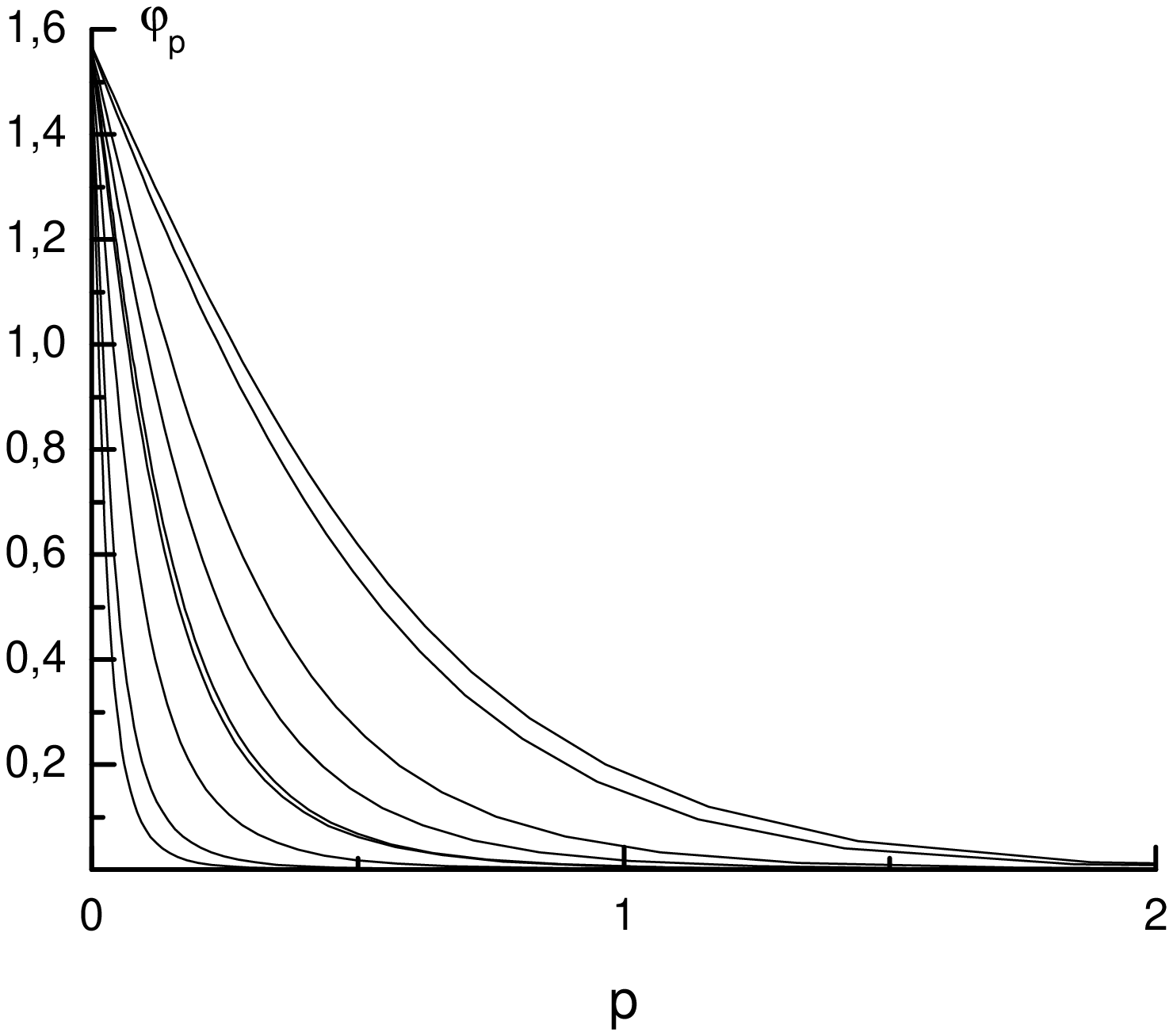,width=7.5cm}\hspace*{5mm}\epsfig{file=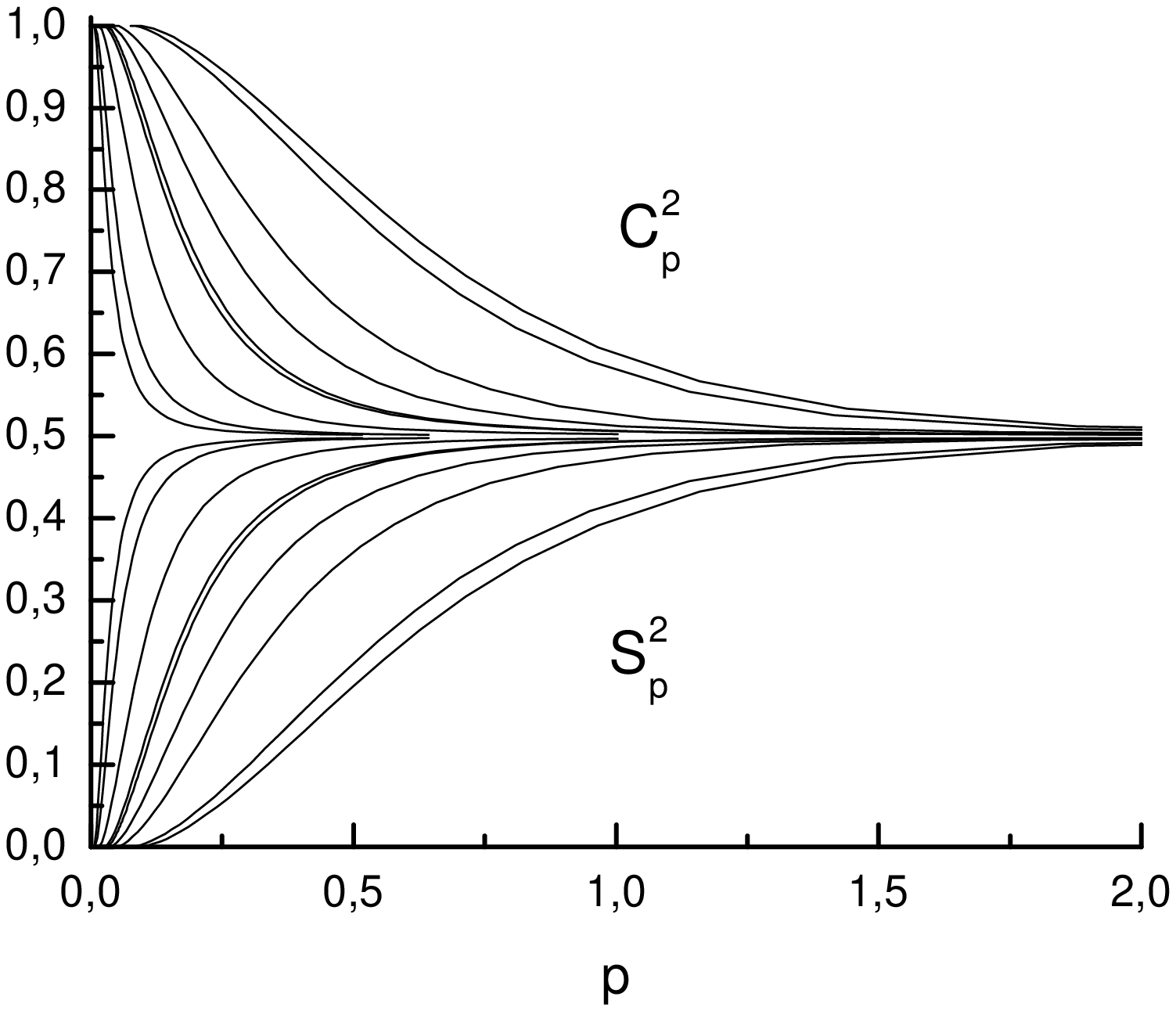,width=7.5cm}}
\caption{Solutions to the mass--gap Eq.~(\ref{mge}) (the first
plot) and the coefficients $C_p^2$ and $S_p^2$ (the second plot)
for the potential $V(r)=K_0^{\alpha+1}r^\alpha$ with various
$\alpha$'s. We plot the curves for
$\alpha=0.3$, 0.5, 0.7, 0.9, 1.0, 1.1, 1.3, 1.7, and 2.0
(smaller $\alpha$'s correspond to steeper curves at the
origin). The momentum $p$ is given in units of $K_0$.}
\end{figure}

For large interquark momenta, the functions $C_p$ and $S_p$ are
smooth and slow, remaining almost constant when approaching their
large-$p$ asymptote of $\frac{1}{\sqrt{2}}$. Therefore, since, for
confining potentials, the distribution given by the Fourier
transform $V(\vpp-\vk)$ has the support at $\vk\approx\vpp$, it is
possible to consider, approximately, $C_pC_k\approx C_p^2$,
$S_pS_k\approx S_p^2$ and to study $C_p^2$ and $S_p^2$ (see the
second plot at Fig.~1). This approximation can be easily checked
for the harmonic oscillator potential, for which
$V(\vpp-\vk)\propto\Delta_k\delta^{(3)}(\vpp-\vk)$. Taking the
corresponding integral by parts, one can omit, for large momenta,
the terms containing derivatives of the almost constant functions
$C_k$ and $S_k$. Thus, the chiral symmetry is restored as soon as
the difference $C_p^2-S_p^2$ becomes negligible. Using the
definition (\ref{CSdef}), one can find that
$C_p^2-S_p^2=\sin\vp_p=\vp_\pi^\pm(p)$, with $\vp_\pi^\pm(p)$
being the wave function of the chiral pion, with coinciding
positive-- and negative--energy components
\cite{Orsay,Orsay2,Lisbon}. Therefore, we conclude that {\em
chiral symmetry is restored in the spectrum if the pionic wave
function vanishes for the given values of the relative interquark
momentum.} This gives us an explicit physically transparent
relation between the BCS scale and the scale of the symmetry
restoration. Although, formally, both scales $\Lambda_{\rm rest}$
and $\Lambda_{\rm BCS}$ are defined by the only dimensional parameter
in the theory $K_0$, we expect, due to numerical factors, the
relation $\Lambda_{\rm rest}\gg\Lambda_{\rm BCS}$ to hold. Below, we
perform a numerical check of this relation and estimate the value
of the restoration scale.

\subsubsection{Numerical estimates: Harmonic oscillator potential}

In this paragraph and in order to illustrate the general results
obtained so far, we perform an explicit numerical study of the
heavy--light mesonic spectrum. To avoid unnecessary complications,
we consider the case of the harmonic oscillator potential, which
allows us to formulate the mass--gap equation and the bound--state
problem through differential equations. This corresponds to the
case $\alpha=2$ for the potential (\ref{potential}). Then the
Fourier transform of the potential is the Laplacian of the
three--dimensional delta--function, 
\be
V(\vpp-\vk)=-K_0^3\Delta_k\delta^{(3)}(\vpp-\vk), 
\ee 
the chiral
angle is the solution to the differential mass--gap equation, 
\be
p^3\sin\vp_p=\frac12K_0^3\left[p^2\vp''_p+2p\vp_p'+\sin2\vp_p\right]+mp^2\cos\vp_p,
\label{diffmge} 
\ee 
and the dressed quark dispersive law can be calculated as 
\be
E_p=m\sin\vp_p+p\cos\vp_p-K_0^3\left[\frac{(\vp'_p)^2}{2}
+\frac{\cos^2\vp_p}{p^2}\right]. 
\label{Epharm} 
\ee 
In order to introduce the BCS scale we, following the approach suggested
before, evaluate the chiral condensate, 
\be
\langle\bar{q}q\rangle=-\frac{3}{\pi^2}\int_0^\infty
dp\;p^2\sin\vp_p\approx -(0.51K_0)^3\equiv -(\Lambda_{\rm BCS})^3,
\label{qaq2} 
\ee 
which gives us 
\be 
K_0=490MeV, 
\label{lc} 
\ee 
if the standard value of the chiral condensate of $-(250MeV)^3$ is
used in order to fix $\Lambda_{\rm BCS}$. Following the definition of
the symmetry restoration formulated above, we demand that the mass
splitting in a chiral doublet $\Delta E$ should be much less than
$\Lambda_{\rm BCS}$, being, at most, just a few per cent of $250MeV$.

In order to solve the bound--state Eq.~(\ref{FW4}), we use, as a
conventional scheme for the quantum numbers $\nu$, the angular
momentum $l$ and the total momentum $j=l\pm\frac12$ of the light
quark, so that the wave function $\psi(\vpp)$ can be expanded as
\be 
\psi(\vpp)=\Omega_{jlm}(\hat{\vec{p}})\frac{u(p)}{p}, 
\ee 
with $u(p)$ being the radial wave function, and the spherical spinors
are defined as 
\be
\Omega_{jlm}(\hat{\vec{p}})=\sum_{\mu_1\mu_1}C^{jm}_{l\mu_1\frac12\mu_2}
Y_{l\mu_1}(\hat{\vec{p}})\chi_{\mu_2}. 
\ee

Then, the eigenvalue equation for the radial wave function $u(p)$
can be readily derived from the bound--state Eq.~(\ref{FW4}) in
the form: 
\be
u''=[E_p-E]u+K_0^3\left[\frac{(\varphi_p')^2}{4}+\frac{(j+1/2)^2}{p^2}+
\frac{\kappa}{p^2}\sin\varphi_p\right]u,
\label{udd} 
\ee 
where
$$
\kappa=\left\{
\begin{array}{ccc}
l,&{\rm for}&j=l-\frac12\\
-(l+1),&{\rm for}&j=l+\frac12
\end{array}
=\pm\left(j+\frac12\right).
\right.
$$
Notice that Eq.~(\ref{udd}) can be rewritten in the form of a Schr{\" o}digerlike equation,
\be
-K_0^3u''+V_{[j,l]}(p)u=Eu,
\label{Se}
\ee
with the effective potential
\be
V_{[j,l]}(p)=E_p+K_0^3\left[\frac14\vp_p^{\prime 2}+\frac{(j+1/2)^2}{p^2}+\frac{\kappa}{p^2}\sin\vp_p\right].
\label{Vnu}
\ee

The well--known property of the spherical spinors, 
\be 
({\vec \sigma}\hat{\vpp})\Omega_{jlm}(\hat{\vec{p}})=-\Omega_{jl'm}(\hat{\vec{p}}),\quad
l+l'=2j, 
\ee 
ensures that the states with $j=l\pm\frac12$ possess
opposite parity. The splitting between such states is the subject
of our investigation, and it follows immediately from
Eq.~(\ref{Vnu}) that, for the given total momentum $j$, this
splitting is due to the $\kappa$--dependent term in this effective
potential. Thus, for $\kappa=\pm(j+\frac12)$, one finds for the
difference of the opposite--parity potentials: 
\be 
\Delta V=-\frac{(2j+1)K_0^3}{p^2}\sin\vp_p, 
\label{DV} 
\ee 
which gives us an explicit example of the relation between the pionic wave
function $\sin\vp_p$ and the opposite--parity state splitting,
discussed above in general terms.

\begin{table}[t]
\caption{The masses of orbitally excited states and the splittings
for the radial quantum number $n=0$, as solutions to the
bound--state Eq.~(\ref{Se}) with the potential (\ref{Vnu}). The
solutions of the Salpeter Eq.~(\ref{Salp0}) are also listed for
the sake of comparison. All energies are given in units of $K_0$.}
\begin{ruledtabular}
\begin{tabular}{ccccc}
$j$&1/2&3/2&5/2&7/2\\
\hline
$E_{l=j-\frac12}$&2.04&3.51&4.51&5.35\\
$E_{l=j+\frac12}$&2.66&3.69&4.57&5.36\\
$\Delta E_j$&0.62&0.18&0.06&0.01\\
\hline
$E_{l=j-\frac12}^{\rm Salp}$&2.34&3.36&4.24&5.05\\
$E_{l=j+\frac12}^{\rm Salp}$&3.36&4.24&5.05&5.79\\
$\Delta E_j^{\rm Salp}$&1.02&0.88&0.81&0.74\\
\end{tabular}
\end{ruledtabular}
\end{table}

\begin{table}[t]
\caption{The masses of orbitally excited states and the splittings
for the radial quantum number $n=1$, as solutions to the
bound--state Eq.~(\ref{Se}) with the potential (\ref{Vnu}). The
solutions of the Salpeter Eq.~(\ref{Salp0}) are also listed for
the sake of comparison. All energies are given in units of $K_0$.}
\begin{ruledtabular}
\begin{tabular}{ccccc}
$j$&1/2&3/2&5/2&7/2\\
\hline
$E_{l=j-\frac12}$&3.91&5.03&5.87&6.60\\
$E_{l=j+\frac12}$&4.39&5.17&5.92&6.61\\
$\Delta E_j$&0.48&0.14&0.05&0.01\\
\hline
$E_{l=j-\frac12}^{\rm Salp}$&4.09&4.88&5.63&6.33\\
$E_{l=j+\frac12}^{\rm Salp}$&4.88&5.63&6.33&7.00\\
$\Delta E_j^{\rm Salp}$&0.79&0.75&0.70&0.67\\
\end{tabular}
\end{ruledtabular}
\end{table}

It is clear from Eq.~(\ref{DV}) and from the form of the chiral
angle (the first plot at Fig.~1) that the splitting vanishes for
excited states, since the excited wave function is localised at
larger relative momenta, whereas the chiral angle decreases fast
with $p$. In order to prove this property explicitly, we solve the
bound--state Eq.~(\ref{udd}) numerically. The results are listed
in Tables~I,II. They demonstrate a clear pattern of chiral
symmetry restoration for orbitally excited heavy--light mesons.
For the sake of comparison, we calculate also the spectrum of the
Salpeter equation 
\be
[\sqrt{\vpp^2+m^2}+K_0^3\vx^2]\psi(\vec{x})=E\psi(\vec{x}),
\label{Salp0} 
\ee 
which, in momentum space, also has the form of
Eq.~(\ref{Se}) with the potential (\ref{Vnu}), with the quark
dispersive law substituted by the free energy and the chiral angle
put equal to $\frac{\pi}{2}$ in the interaction, which is, 
\be
V_{[j,l]}^{\rm Salp}(p)=\sqrt{p^2+m^2}+K_0^3\frac{(j+1/2)^2+\kappa}{p^2}=\sqrt{p^2+m^2}+K_0^3\frac{l(l+1)}{p^2}.
\ee 
The results are also listed in Tables~I,II.

\begin{figure}[t]
\centerline{\epsfig{file=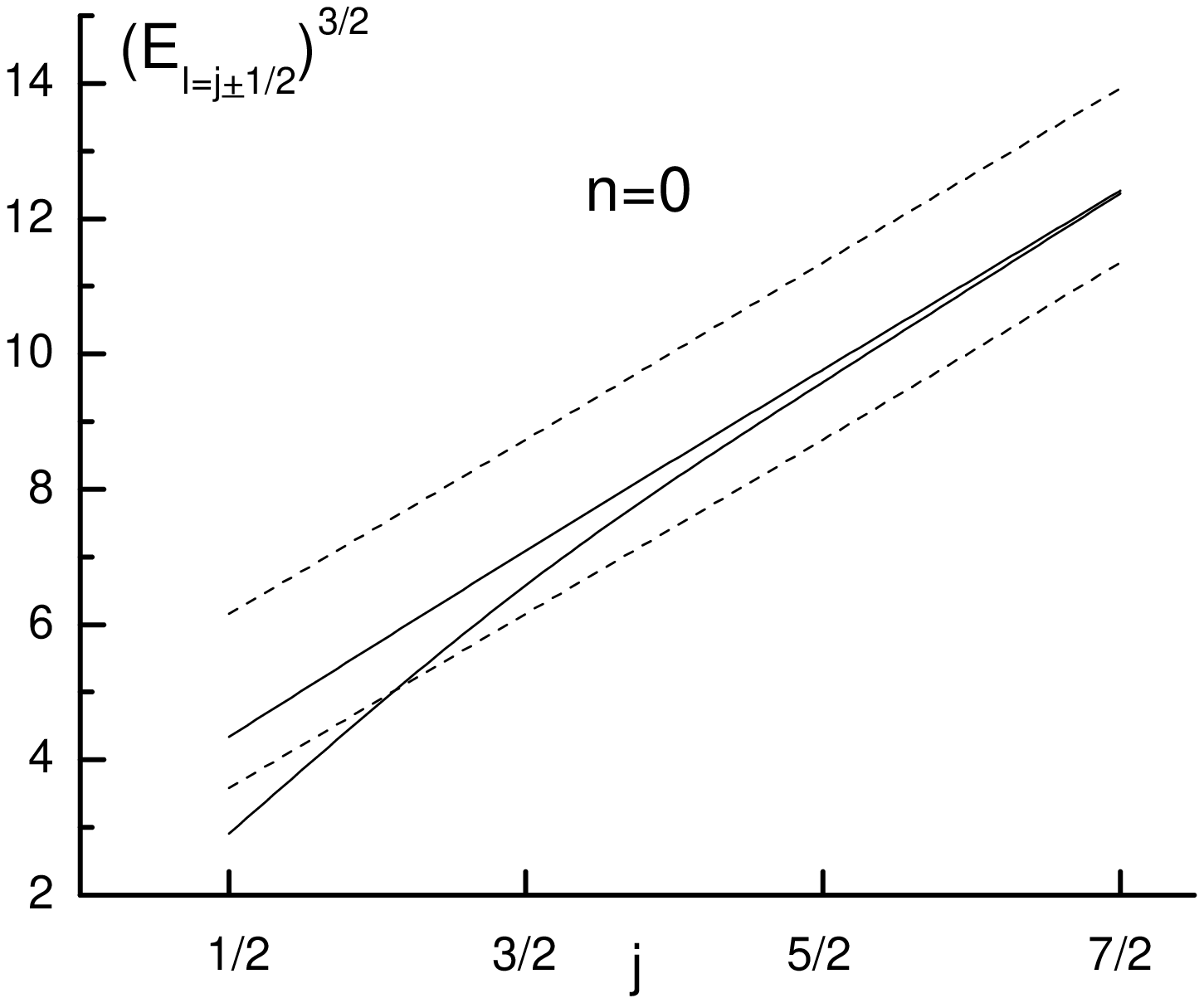,width=7cm}\hspace*{10mm}\epsfig{file=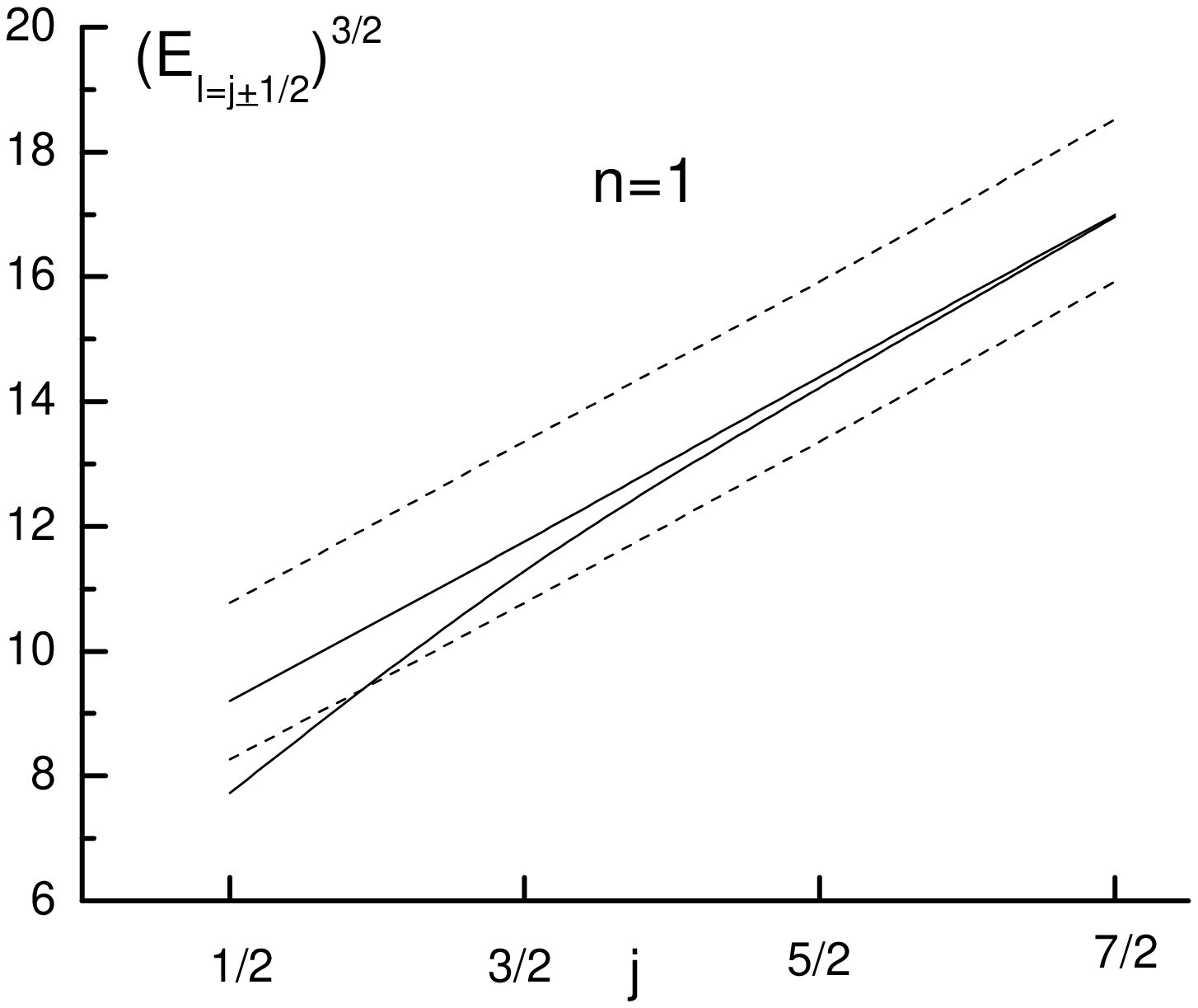,width=7cm}}
\caption{Regge trajectories for Eq.~(\ref{Se}) with the potential
(\ref{Vnu}) (solid lines) and for the Salpeter Eq.~(\ref{Salp0})
(dashed lines). In each pair, the lower curve corresponds to
$l=j-\frac12$ and the upper curve corresponds to $l=j+\frac12$.}
\end{figure}

In Fig.~2 we plot the Regge trajectories for the bound--state
Eqs.~(\ref{Se}) and (\ref{Salp0}), as $E^{3/2}$ versus
$j$\ftnote{1}{The quasiclassical spectrum of Eq.~(\ref{Salp0})
gives $E\propto l^{2/3}$. Therefore, for this equation, for the
given radial excitation number $n$ and the total momentum $j$, one
expects two neighbouring nearly parallel straight line
trajectories $E^{3/2}(j)$, corresponding to
$l=j\pm\frac12$. Fig.~2 clearly exhibits such a behaviour.}, for
the radial quantum numbers $n=0$ (the first plot) and $n=1$ (the
second plot). For the bound--state Eq.~(\ref{Se}), the
trajectories for $j=l\pm\frac12$ merge in such a way that at
$j=5/2$ the splitting is $0.06$ (in the units of $K_0$), whereas
for $j=7/2$ it is already $0.01$. Now the BCS scale
$\Lambda_{\rm BCS}$ is $0.51 K_0$. Therefore, if we require that the
splitting constitutes a few per cent of $0.51$ (say, about 10
per cent), then the splitting should be below $0.051$, which
happens between $j=5/2$ and $j=7/2$, at $E^{3/2}\simeq 11$. This
gives the bound--state energy $E\approx 5 K_0$. 
Let us define this energy to be the restoration scale $\Lambda_{\rm rest}$.
For $j=7/2$ we have a mass splitting which, despite being already
fifty times smaller than $\Lambda_{\rm BCS}$, yields a bound--state energy
(of $\approx 5.3$) just slightly above $\Lambda_{\rm rest}$. It is
then clear that $\Lambda_{\rm rest}$ can be thought of as the
onset scale for the chiral restoration regime. 
In other words: for a
mass splitting 10 times smaller than the BCS scale
$\Lambda_{\rm BCS}$ we have $\Lambda_{\rm rest}$ 10 times bigger than
the BCS scale $\Lambda_{\rm BCS}$, 
\be 
\Lambda_{\rm rest}\approx 5K_0\approx 2.5GeV\approx 10 \Lambda_{\rm BCS}, 
\label{rel} 
\ee 
where the estimate (\ref{lc}) was used for the parameter $K_0$. 
The result (\ref{rel}) is in good agreement with other estimates of the
restoration scale known in the literature \cite{Swanson}. 
In Fig.~3 we illustrate, for the sake of completeness, the behaviour of the splitting $\Delta E_j=E_{l=j+\frac12}-E_{l=j-\frac12}$ 
as a function of the averaged bound--state energy $E_j=\frac12(E_{l=j+\frac12}+E_{l=j-\frac12})$.

\begin{figure}[t]
\centerline{\epsfig{file=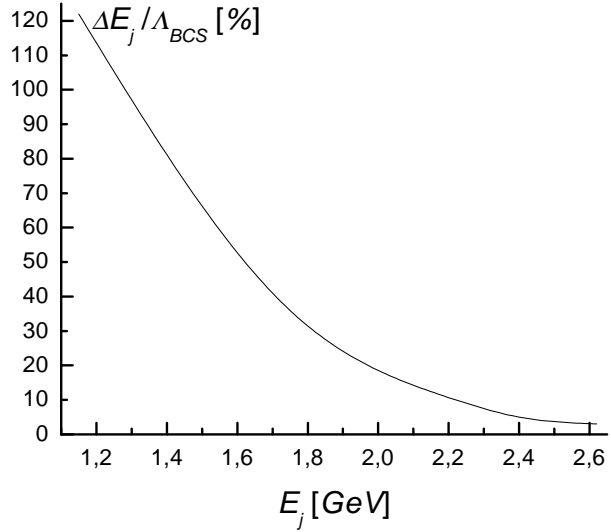,width=8cm}}
\caption{The ratio of the splitting $\Delta E_j=E_{l=j+\frac12}-E_{l=j-\frac12}$ to the BCS scale $\Lambda_{\rm BCS}$, 
in per cent, versus the averaged bound--state mass $E_j=\frac12\left(E_{l=j+\frac12}+E_{l=j-\frac12}\right)$. 
For the sake of transparency we measure the bound--state energy in $GeV$, using the estimate (\ref{lc}).}
\end{figure}

In contrast, the trajectories for Eq.~(\ref{Salp0}) remain
parallel for all values of the total momentum $j$. This contrast
provides an explicit demonstration of the role played for the
restoration of the symmetry by the pionic wave function
$\sin\vp_p$ in the potential (\ref{Vnu}).

\subsubsection{Generalisation to a generic form of the potential}

The results obtained in the framework of the potential quark model
(\ref{H}) are known to be robust against variations of the form of
the quark kernel --- although the values of various physical
quantities might change, for example, with the change of the shape
of the confining potential, the relations between such quantities
are left intact. 

In Fig.~4, for the sake of clarity, we plot the coefficients $C_p^2$  and $S_p^2$ versus the relative momentum 
$p$, measured in $GeV$. To this end we extracted the corresponding scale $K_0$, for the given power $\alpha$, from the 
definition of the BCS scale --- see Eqs.~(\ref{qaq2}), (\ref{lc}). This plot is to be compared with the second plot in Fig.~1, where the
same quantities are plotted in the units of $K_0$. One can see the universality of the behaviour of these functions,
regardless of the explicit from of the confining potential. A similar conclusion holds for the dressed quark dispersive law $E_p$ which, for
excited states in the spectrum, tends to the $\alpha$--independent free--quark limit of $\sqrt{p^2+m^2}$. We conclude, therefore, that the
only ingredient of the bound--state Eq.~(\ref{FW4}) which changes with the change of the confining potential is the Fourier transform
$V(\vpp-\vk)$, the net result of this change being simply the modification of the Regge trajectory form, from $E^{3/2}$ {\em vs} $j$, for the
harmonic oscillator potential, for $E^{\frac{\alpha+1}{\alpha}}$ {\em vs} $j$, for the generic potential (\ref{potential}). 
Therefore, we dare to extend the relation
(\ref{rel}) beyond the harmonic oscillator potential case of
$\alpha=2$, generalising it to the entire family of allowed
confining potentials. We find it remarkable that the chiral
potential model suggested for studies of chiral low--energy
phenomena in QCD appears able to address the problem of the chiral
symmetry restoration in the spectrum of highly excited hadrons and
to give a reasonable robust prediction for the corresponding
restoration scale.

\begin{figure}[t]
\centerline{\epsfig{file=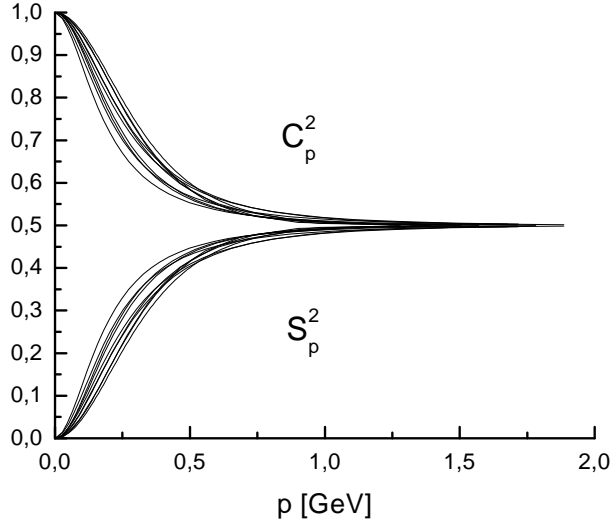,width=8cm}}
\caption{The coefficients $C_p^2$ and $S_p^2$ for the potential $V(r)=K_0^{\alpha+1}r^\alpha$ with various
$\alpha$'s (we use the same set of $\alpha$'s as in Fig.~1).}
\end{figure}

\section{Conclusions}

In this paper, we investigate the chiral symmetry restoration for
highly excited states in the hadronic spectrum. We consider the
NJL-type model (\ref{H}), with an arbitrary confining quark
kernel, and use it to study the heavy--light quarkonium. We derive
the bound--state equation for the given system in the form of a
Schr{\" o}dingerlike equation and also in the form of an
effective Diraclike equation for the light quark in the field of
the static antiquark. We give an explicit expression for the
matrix structure of the effective interaction of the light quark
with the antiquark source for all relative interquark momenta and
study the Lorentz nature of this interaction. Thus, we identify
explicitly the following three regimes: i) the chiral regime,
$p\lesssim\Lambda_{\rm BCS}$, with chiral symmetry breaking playing a
dominating role in the interaction --- the latter being
predominantly scalar, ii) the restoration regime, $\Lambda_{\rm
rest}\lesssim p$, which realises for highly excited states in the
spectrum, with the effective interaction being vectorial and, as a
result, states with opposite parity coming in doublets, and iii)
the intermediate regime, which interpolates the two aforementioned
ones. We perform a numerical investigation of the heavy--light
bound--state equation for the harmonic oscillator potential and
demonstrate explicitly the effect of chiral symmetry restoration
in the spectrum for orbitally excited states. We estimate the
restoration scale to be $\Lambda_{\rm rest}\approx
10\Lambda_{\rm BCS}\approx 2.5GeV$, where the BCS scale
$\Lambda_{\rm BCS}$, defining the low--energy properties of the theory,
for example, related to the chiral condensate, is chosen to take
the standard value of about $250MeV$. Bearing in mind robustness
of predictions of the theory (\ref{H}) for relations between
various physical parameters with respect to variations of the
quark kernel, we dare to extend this conclusion to an arbitrary
confining interquark kernel in the Hamiltonian (\ref{H}). Thus,
we conclude that there is a sufficient window for the intermediate
regime in which, on one hand, chiral symmetry does not play a
dominating role anymore, whereas, on the other hand, the parity
doubling still does not happen in the spectrum. We find that the
scale of the chiral symmetry restoration in the spectrum,
evaluated in the framework of the potential models (\ref{H}), is
in good agreement with other estimates known in the literature.

\begin{acknowledgments}
Yulia Kalashnikova and Alexei Nefediev acknowledge the financial support of the grant
NS-1774.2003.2, as well as of the Federal Programme of the Russian
Ministry of Industry, Science, and Technology No 40.052.1.1.1112.
Emilio Ribeiro acknowledges Enrico Maglione for useful discussions
on the issue of numerical evaluation of bound states and Alexei Nefediev
would like to thank Yu. A. Simonov for stimulating discussions.
\end{acknowledgments}

\end{document}